\documentclass[conference]{IEEEtran}
\usepackage{makecell}
\IEEEoverridecommandlockouts
\usepackage[linesnumbered,ruled,vlined]{algorithm2e}
\usepackage{booktabs}
\usepackage{multirow}
\usepackage{cite}
\usepackage{amsmath,amssymb,amsfonts}
\usepackage{algorithmic}
\usepackage{graphicx}
\usepackage{textcomp}
\usepackage{xcolor}
\usepackage{comment}
\usepackage{geometry}
\usepackage{hyperref}

\usepackage{float}

\usepackage{booktabs} 
\usepackage{array}    
\begin{document}
\title{PRISM: Privacy-preserving Inference System with Homomorphic Encryption and Modular Activation}

\author{
    \IEEEauthorblockN{
        Zeinab Elkhatib$^{1}$, Ali Sekmen $^{1}$, Kamrul Hasan$^{1}$}
    \IEEEauthorblockA{
        $^{1}$ Tennessee State University, Nashville, TN, USA \\
        Email: $\lbrace$\textit{zelkhati, asekmen, mhasan1}$\rbrace$@tnstate.edu
    }
}
\maketitle

\begin{abstract}
With the rapid advancements in machine learning, models have become increasingly capable of learning and making predictions in various industries. However, deploying these models in critical infrastructures presents a major challenge, as concerns about data privacy prevent unrestricted data sharing. Homomorphic encryption (HE) offers a solution by enabling computations on encrypted data, but it remains incompatible with machine learning models like convolutional neural networks (CNNs), due to their reliance on non-linear activation functions. To bridge this gap, this work proposes an optimized framework that replaces standard non-linear functions with homomorphically compatible approximations, ensuring secure computations while minimizing computational overhead. The proposed approach restructures the CNN architecture and introduces an efficient activation function approximation method to mitigate the performance trade-offs introduced by encryption. Experiments on CIFAR-10 achieve 94.4\% accuracy with 2.42~s per single encrypted sample and 24,000~s per 10,000 encrypted samples,  using a degree-4 polynomial and Softplus activation under CKKS, balancing accuracy and privacy.

\end{abstract}

\section{Introduction}
Neural networks have demonstrated remarkable predictive performance and delivered innovative solutions across a broad spectrum of domains. However, when deploying these models in sensitive areas such as healthcare and critical infrastructures, it is imperative that they comply with strict data protection regulations, including HIPAA \cite{hipaa1996} and GDPR \cite{gdpr2016}. A promising strategy to address these challenges is to use publicly available data for training while employing homomorphic encryption during inference. Training on non-sensitive, public datasets minimizes the risks associated with data alteration or exposure during model development, allowing us to focus our security efforts on the inference phase \cite{leejon}. This approach ensures that sensitive data remains encrypted throughout the computational process, thereby providing a robust and compliant framework for real-world deployment.

In the inference phase, applying Homomorphic encryption on the confidential data offers a powerful means to preserve data confidentiality by enabling computations to be performed directly on encrypted data. In this context, the CKKS fully homomorphic encryption scheme, named after its developers Cheon, Kim, Kim, and Song, is particularly attractive. CKKS is specifically designed for approximate arithmetic on ciphertexts, making it feasible to perform neural network inferencing without exposing the underlying sensitive information \cite{zhang2021galagreedycomputationlinear}.
Nonetheless, integrating CKKS with convolutional neural networks, CNNs, presents significant challenges. A primary obstacle is the reliance of CNNs on non-linear activation functions, such as ReLU, Sigmoid, Tanh, and Swish, which are not directly compatible with the algebraic operations supported by CKKS \cite{obla}. To bridge this gap, researchers have developed polynomial approximations of these activation functions. These techniques can be broadly categorized into two groups. The first category utilizes lower-degree polynomials with a fixed number of multiplications, thereby prioritizing computational efficiency at the cost of reduced model accuracy \cite{albadawi}, \cite{herve,chou2018}. In contrast, the second category employs higher-degree approximations in conjunction with bootstrapping to mitigate the noise introduced in ciphertexts, particularly in deeper network architectures. Although this approach can achieve higher inference accuracy, it comes with significant computational overhead and increased resource consumption \cite{sinem}.
Moreover, the challenge extends beyond activation functions. Other components of the CNN architecture, including convolution and pooling operations, also require adjustments to operate efficiently under the constraints of CKKS \cite{herve}. These modifications are necessary to ensure that all parts of the network are compatible with encrypted computation.
The trade-off between computational efficiency and inference accuracy raises a critical question: Is it possible to design a convolutional neural network architecture with a simplified activation function that achieves high accuracy while minimizing computational time and resource usage?
In this work, we address this question by proposing a novel design that carefully balances these competing factors. Our approach leverages a low-degree polynomial approximation for activation functions and modifies the CNN architecture to better suit encrypted computations. Furthermore, we focus exclusively on secure inference using homomorphic encryption while relying on publicly available datasets for training. This strategy offers a practical and efficient solution for applications in critical infrastructures, where data confidentiality, regulatory compliance, and real-time performance are essential.\\
\noindent The rest of this paper is organized as follows: Section II reviews related work; Section III presents the methodology; Section IV discusses the experimental results; and Section V concludes the study.

\section{Related Work}

In this section, previous works can be seen to have developed polynomial approximations and adjusted the Convolutional Neural Network (CNN) architecture to become compatible with handling images in the form of ciphertext. The results of each work can be seen in Table I and Table II representing the MNIST and CIFAR-10 datasets, respectively. It can be seen that for MNIST, works \cite{herve, albadawi, highly}, have similar accuracies and inference per image time. And even when increasing the degree of the polynomial \cite{highly}, the accuracy stays relatively similar to the rest. This is because the MNIST dataset does not have too many features. Because of the limited number of features, using a shallow CNN architecture, employing a simple polynomial approximation, such as using the square function as in \cite{herve}, \cite{albadawi} can yield surprisingly good results. This is because a shallow network, by design, minimizes the number of layers and operations, thereby reducing both computational overhead and cumulative noise in homomorphic encryption. As a result, even basic approximations are sufficient to capture the necessary nonlinearities for accurate inference while keeping the evaluation process efficient and within the noise budget.
In homomorphic encryption for CNNs on datasets like CIFAR 10, striking the right balance between network depth and the degree of polynomial approximation is crucial. CIFAR 10 images,32×32×3, \cite{abouelnaga2016cifar} are complex and require an architecture capable of extracting detailed, hierarchical features. A shallow CNN with only a few layers may reduce computational costs, and a simple low-degree polynomial (such as the square function) can offer a computationally efficient approximation. However, this combination, while efficient, tends to yield lower accuracy (for instance, 77.59\% as shown in Table 2 \cite{albadawi}) because the network lacks the depth to capture all the nuances present in the data. Increasing the polynomial degree to between 15 and 27 \cite{leejon} does improve accuracy, yet it still may not match the performance of more sophisticated CNN designs like those in \cite{obla}, which achieved around 90\%+ accuracy using a lower degree polynomial by employing a more effective architecture. On the flip side, deeper CNN architectures introduce a higher number of homomorphic operations, which in turn accumulates noise in the ciphertexts. When this noise grows too high, decryption fails unless a process called bootstrapping is applied. Bootstrapping “refreshes” the ciphertexts by reducing accumulated noise, but it is extremely time-consuming, sometimes extending inference times to as long as three hours per image, as observed in \cite{leejunghyun}. Therefore, while a shallow network with a simple polynomial is computationally attractive, it often cannot extract the rich features needed from CIFAR 10, and a deeper network, despite its potential for higher accuracy, may become impractical due to the heavy computational burden and the need for frequent bootstrapping. The key challenge is finding a middle ground that offers sufficient expressive power without incurring prohibitive computational delays.

\begin{table}[ht]
  \centering
  \caption{Results of Previous Works on Mnist Dataset}
  \label{tab:results1}
  \begin{tabular}{l *{5}{c}}
    \toprule
    Author & \multicolumn{4}{c}{Results: MNIST} \\
    \cmidrule(lr){2-5}
           & Activation & Accuracy & Time & Degree \\
    \midrule
    Chabanne et al. \cite{herve}.  & Square & 99.3\%   & 2.58 ms & Degree 2 \\ 
    Badawi et al. \cite{albadawi} & Square   & 99.0\%   & 2.75 ms & Degree 2 \\
    Srinath et al.\cite{obla} & Softplus    & 99.59\%   & 2.50 ms & Degree 2 \\ 
    Takumi et al.\cite{highly} & Swish & 99.29\%   & 21.15s & Degree 5 \\ 
    Hesamifard et al.\cite{Hesamifard2017CryptoDL} & ReLU    & 99.25\%   & 2.55 ms & Degree 3 \\
    
    \bottomrule
  \end{tabular}
\end{table}

\begin{table}[ht]
  \centering
  \caption{Results of Previous Works on Cifar-10 Dataset}
  \label{tab:results2}
  \begin{tabular}{l *{4}{c}}
    \toprule
    Author & \multicolumn{4}{c}{Results: CIFAR-10} \\
    \cmidrule(lr){2-5}
           & Activation & Accuracy & Time & Degree \\
    \midrule
    Proposed Method & Softplus & 89.65\% & 17.38s & Degree 4 \\ 
    Badawi et al.\cite{albadawi}  & Square   & 77.59\% & 304.43s  & Degree 2 \\
    Srinath et al.\cite{obla} & Softplus & 90.37\% & --       & Degree 7 \\ 
    Takumi et al.\cite{highly}  & ReLU     & 81.06\% & 1555.5s  & Degree 4 \\ 
    Junghyun et al.\cite{leejon}& ReLU     & 87.9\%  & 2,892s   & Degree 15--27 \\ 
    Joon-Woo et al.\cite{leejunghyun}& ReLU     & 92.43\% & 3hr      & Degree 16 \\ 
    Chou et al.\cite{chou2018}    & Swish    & 75.99\% & 22,372s  & Degree 2 \\
    \bottomrule
  \end{tabular}
\end{table}

\section{Methodology}

To improve the classification accuracy and computational efficiency of convolutional neural networks (CNNs) evaluated under homomorphic encryption (HE), we propose a complete methodology for training and deploying encrypted-compatible models. The primary constraint in encrypted inference include the need to replace the non-linear activation function with its polynomial equivalent. This is done to be compatible with encypted data and to minimize multiplicative depth in order to preserve ciphertext integrity and reduce evaluation time. To address these challenges, we approximate the Softplus function using a low-degree polynomial generated through weighted minimax optimization using Powell’s method. Next, we applied batch normalization (BN) to constrain input distributions and improve approximation accuracy. Furthermore, we modified the CNN architecture to eliminate non-polynomial operations while preserving the model's depth. Finally, we train models using a two-phase procedure which allows the model to learn the weights of the model and implement them during the inference stage. The experiments were conducted on a system equipped with an Intel\textsuperscript{\textregistered} Core\textsuperscript{TM} i9-14900HX CPU (24 cores, 32 threads, base clock 2.20~GHz) and 31.71~GB of RAM, with evaluation focused on classification accuracy and inference time. The entire model can be visually seen in the figure below.

\begin{figure*}[h!]
    \centering
    \includegraphics[width=0.8\textwidth,height=0.30\textheight]{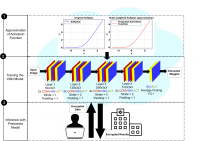}
    \caption{Complete Model Described}
    \label{fig:relu}
\end{figure*}

\subsection {Selection of Activation Function}
In this work, we propose a HE- friendly activation function by using the Softplus function. It is defined as: \begin{equation}
\text{Softplus}(x) = \log\left(1 + e^x\right)
\end{equation}
and is approximated into a degree-4 polynomial in the form below: \begin{equation}
f(x) = A x^4 + B x^3 + C x^2 + D x + E
\end{equation}.

Softplus was selected because it exhibits a gradual transition and produces smoother curves, making it well-suited for approximation with a lower-degree polynomial. This is particularly important for homomorphic encryption, where lower-degree polynomial approximations reduce the number of multiplicative operations, thereby limiting the multiplicative depth. A lower multiplicative depth is critical because it minimizes noise accumulation, which can otherwise degrade evaluation accuracy. 

Furthermore, although ReLU and Swish are popular activation functions and often perform well in CNNs on the CIFAR 10 dataset, our approximation experiments show that both produce higher maximum approximation errors than Softplus, as shown in Table~\ref{tab3}, when subject to the same degree constraint.

\begin{table}[h!]
\caption{Polynomial approximation details and max errors for different activation functions.}
\centering
\label{tab3}
\begin{tabular}{lccc}
\toprule
\makecell{\textbf{Activation} \\ \textbf{Function} \\ \textbf{Approximated}} & 
\makecell{\textbf{Degree}} & 
\makecell{\textbf{Range}} & 
\makecell{\textbf{Max Error}} \\
\midrule
Softplus & 4 & [-7, 7] & 0.067 \\
ReLU     & 4 & [-7, 7] & 0.331 \\
Swish    & 4 & [-7, 7] & 0.18 \\
\bottomrule
\end{tabular}
\end{table}

This is expected, especially for ReLU, whose sharp, non-smooth shape makes it harder to approximate with a low-degree polynomial. Swish also shows a larger deviation than Softplus when comparing its analytic and polynomial forms. Because of this, Softplus is a more practical choice for homomorphic settings, even if its raw performance in plaintext CNNs is slightly lower. To fairly compare these functions, we evaluated the quality of each approximation through a three-step process: an initial weighted least-squares fit, Powell’s minimax optimization, and a final maximum-error evaluation. For the Softplus activation, this procedure yielded the degree-4 polynomial with optimized coefficients: $A = -0.00068481$, $B = -1.5983 \times 10^{-17}$, $C = 0.0887234775$, $D = 0.5$, and $E = 0.738099333$.

The approximability of Softplus over this interval can also be justified mathematically. The Softplus function, $\mathrm{Softplus}(x)=\log(1+e^x)$, is analytic on the real line and extends smoothly to the complex plane except for isolated branch points where $1 + e^z = 0$, located at $z = i\pi(2k + 1)$ for integer $k$. Consequently, $\mathrm{Softplus}(x)$ is analytic within the horizontal strip $|\mathrm{Im}(z)| < \pi$. After rescaling $x$ to $t = x/7$ so that $t \in [-1,1]$, the function $g(t)=\mathrm{Softplus}(7t)$ remains analytic in the narrower strip $|\mathrm{Im}(t)| < \pi/7$. Classical results in polynomial approximation theory, such as Bernstein’s and Weierstrass’s theorems, guarantee that the best degree-$n$ polynomial approximation to an analytic function in such a strip converges exponentially with $n$. Specifically, if $g$ is analytic in $|\mathrm{Im}(t)| < \alpha$, then there exists $\rho = e^{\alpha} > 1$ such that the minimax approximation error satisfies $\|g - p_n\|_{\infty,[-1,1]} = \mathcal{O}(\rho^{-n})$. For $\alpha = \pi/7$, we obtain $\rho \approx 1.566$, implying an error reduction proportional to $1.566^{-n}$. For a fourth-degree polynomial ($n=4$), this gives an upper bound of approximately $0.106$, consistent with our empirical minimax error of $E_{\max} = 0.067$. This confirms that Softplus can be effectively approximated with low-degree polynomials due to its smooth, analytic structure.

To begin, the approximation domain was set to $[-7,7]$, chosen after batch-normalizing the CIFAR-10 dataset and observing that approximately $99.7\%$ of activations fell within $[-3,3]$, with a small fraction ($\sim 0.03\%$) extending beyond this range. Expanding the domain to $[-7,7]$ ensures the polynomial remains accurate for both common and rare activation values. Within this domain, region-specific weights were assigned to reflect the relative importance of different subranges during CNN inference. The central region $[-3,3]$, which contributes most heavily to model performance, was assigned a weight of 3, while the secondary ranges $[-7,-4]$ and $[4,7]$ were given moderate weights of 2, and all remaining points were assigned a baseline weight of 1. This weighting scheme emphasizes high-probability activation regions while maintaining global smoothness across the domain.

The initial polynomial was obtained by solving the weighted least-squares problem:
\begin{equation}
\min_{\mathbf{c}} \sum_{i=1}^{N} w(x_i) \left(p_{\mathbf{c}}(x_i) - f(x_i)\right)^2,
\end{equation}
where $w(x_i)$ is the region-specific weight for sample point $x_i$, $f(x_i)$ is the analytic Softplus value, and $p_{\mathbf{c}}(x_i)$ is the polynomial with coefficients $\mathbf{c}$. This formulation ensures that the approximation error is minimized more aggressively in critical regions, producing a smooth, stable, and high-fidelity approximation suitable for encrypted CNN inference.

Powell’s derivative-free optimization algorithm was employed to refine the coefficients obtained from the initial weighted least-squares (WLS) fit. The goal was to minimize the maximum weighted absolute error between the analytic Softplus function $f(x)$ and its polynomial approximation $p_{\mathbf{c}}(x)$ over the interval $[-7,7]$, defined as
\begin{equation}
\min_{\mathbf{c}} \ \max_{x_i \in [-7,7]} w(x_i)\,|f(x_i) - p_{\mathbf{c}}(x_i)|,
\end{equation}
where $\mathbf{c} = [A, B, C, D, E]$ are the polynomial coefficients and $w(x_i)$ are the region-specific weights emphasizing the dense activation range $[-3,3]$. This formulation represents a weighted minimax problem that explicitly minimizes the largest weighted deviation between $f(x)$ and $p(x)$. Since the objective function is non-convex and non-differentiable due to the absolute value and maximum operators, Powell’s method is particularly well suited because it performs successive line minimizations along conjugate search directions without requiring gradients, iteratively updating directions to reduce the worst-case deviation until convergence. The optimization was initialized with the coefficients obtained from the WLS fit, providing a smooth starting point close to the least-squares optimum. Each coefficient was treated as an unconstrained variable in $\mathbb{R}^5$, and no explicit bounds were imposed since the polynomial form remains numerically stable within the bounded domain $[-7,7]$. The algorithm terminated when the reduction in the objective $E_{\max}$ between successive iterations fell below $10^{-8}$ or after 200 iterations, whichever occurred first. Across multiple random initializations, the algorithm consistently converged to identical coefficients (agreement within $10^{-5}$), confirming a stable local minimum that is empirically near-global. This optimization stage was integrated as the final step in the approximation pipeline: the WLS fit generated an initial solution, Powell’s method refined it by minimizing the worst-case deviation, and the resulting coefficients were then evaluated under an unweighted setting to measure the true maximum deviation
\begin{equation}
\varepsilon(x) = |p(x) - f(x)|, \quad E_{\text{max}} = \max_{x \in [-7,7]} \varepsilon(x).
\end{equation}
All optimization steps were performed over $[-7,7]$, selected based on empirical batch-normalized activation distributions where over 99\% of activations fall within $[-3,3]$. This ensures that the polynomial maintains high fidelity in the region most critical during inference. Within this domain, the degree-4 approximation achieved a small worst-case error ($E_{\max} = 0.067$), with deviations primarily near the boundaries. Because batch normalization centers activations around zero, the effective operating region exhibits an even smaller average error. Increasing the polynomial degree to 6–8 reduced $E_{\max}$ by less than 15\% while substantially increasing multiplicative depth, confirming that the degree-4 polynomial offers the best balance between approximation accuracy and homomorphic efficiency for batch-normalized activations. Finally, we verified that the degree-4 Softplus polynomial is the \textit{global minimax optimum} over $[-7,7]$. Using a linear programming epigraph formulation of the weighted minimax problem, we confirmed that no lower maximum error exists on a dense grid. The solution produced a stable maximum weighted error ($E_{\max}=0.1243$) across multiple refinements and exhibited the characteristic equiripple alternation pattern, providing strong numerical evidence of global optimality under the chosen weighting scheme.

\begin{figure}[h!]
    \centering
    \includegraphics[width=0.40\textwidth,height=0.20\textheight]{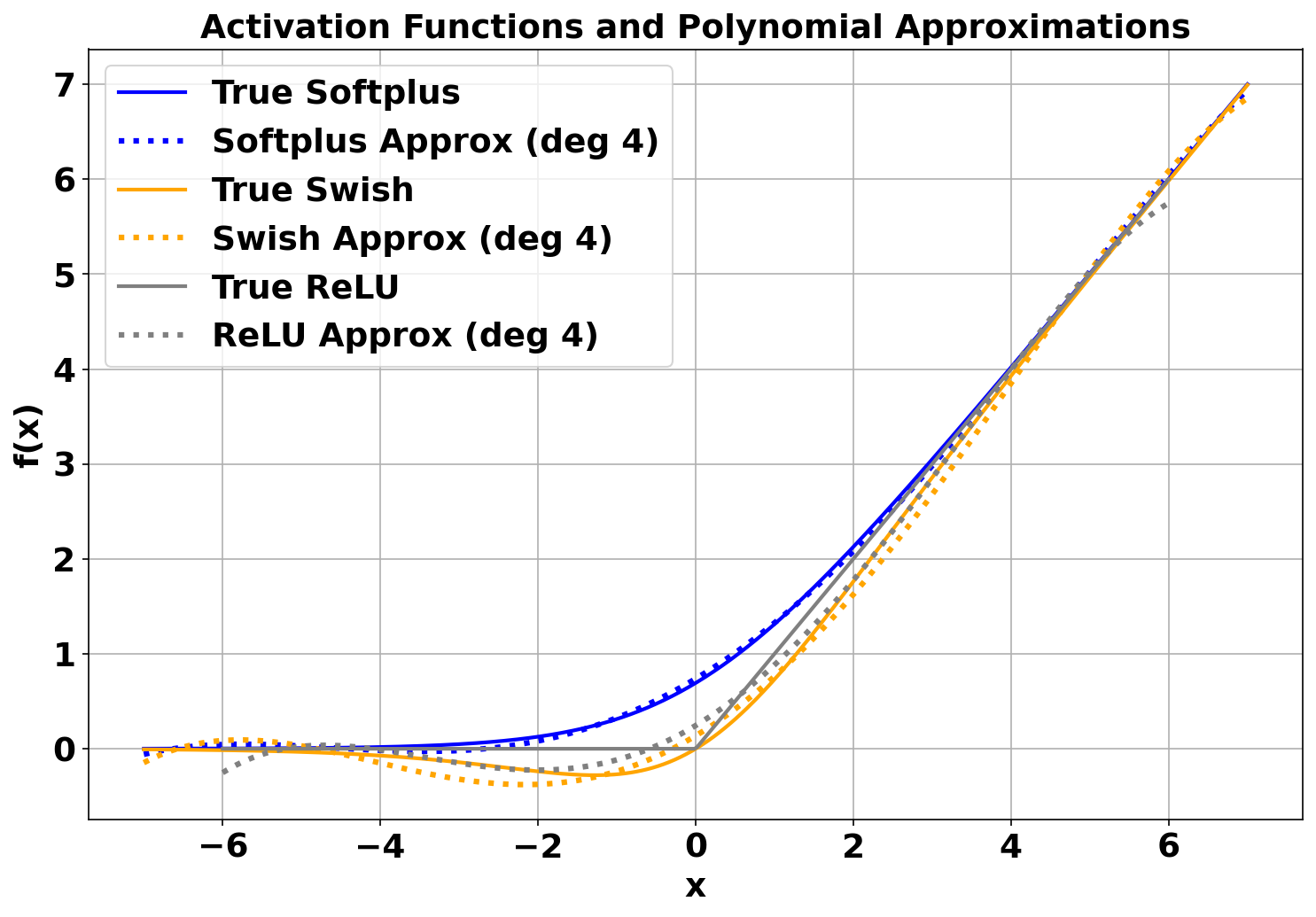}
    \caption{Swish, ReLU, and Softplus Activation Function Approximations}
    \label{fig:activation}
\end{figure}

\subsection{Network Architecture Used in the Training Phase}

To evaluate the proposed activation approximation framework, we implemented a convolutional neural network tailored for encrypted inference on the CIFAR-10 dataset. The architecture follows a residual-style design \cite{he2016identity} but uses a post-activation block structure, where each convolution is directly followed by batch normalization and the proposed degree 4 polynomial Softplus activation. This ordering can be seen in Fig. 1, and it allows batch normalization to be folded into layers for homomorphic evaluation. The network is composed of four convolutional stages. The first stage applies two 3×3 convolutions with 64 channels, stride 1, and padding 1. The second stage increases the channel count to 128 using two 3×3 convolutions with stride 2 and padding 1, which also downsample the feature maps. The third stage continues the same two-convolution pattern, expanding the channel depth to 256 with stride 2. The fourth stage again applies two 3×3 convolutions, this time with 512 channels and stride 2, further reducing the spatial resolution. After these four stages, a final batch normalization and polynomial activation are applied, followed by a global average pooling layer and a fully connected layer mapping to 10 output units, corresponding to the CIFAR-10 classes. In total, the model contains eight convolution, batch normalization and activation blocks, arranged in pairs across the four stages, with progressively increasing channel sizes of 64, 128, 256, and 512, and strides of 1, 2, 2, and 2 respectively. All nonlinearities use the proposed degree-4 polynomial Softplus approximation with input clamping to reduce approximation error. This design eliminates non-HE-friendly components such as max-pooling and ReLU while maintaining strong representational capacity, ensuring compatibility with homomorphic encryption constraints.

\subsection{Two-Stage Framework for Training and Encrypted Inference}

Our methodology employs a two-stage framework designed to enable secure inference under homomorphic encryption (HE) while preserving compatibility with conventional deep learning workflows. In both stages, the network only uses the approximated Softplus activation function, ensuring that the model remains fully HE-compatible from training through deployment. In the first stage, the model described in the 'Network Architecture in the Training Phase' section is trained entirely on plaintext (unencrypted) data. This choice is motivated by two practical reasons. First, it reflects common real-world scenarios in which datasets are already collected and stored in unencrypted form prior to model deployment, making plaintext training both feasible and realistic. Second, performing training without encryption allows us to take advantage of standard GPU acceleration and optimized deep learning libraries, thereby greatly reducing computational costs and training time. Since homomorphic encryption imposes a significant computational overhead, avoiding it during training prevents unnecessary slowdowns and allows for more extensive experimentation and hyperparameter tuning. In the second stage, the trained model is adapted for encrypted inference. This involves replacing plaintext inputs with their CKKS-encrypted counterparts and executing most forward-pass operations in the encrypted domain. By separating training and encrypted inference into distinct phases, we are able to leverage the efficiency and flexibility of raw-data training while confining the computationally intensive HE operations to the inference stage, where privacy preservation is essential. This two-stage design ensures that the model achieves high accuracy and stability during training, while still supporting fully privacy-preserving predictions at deployment, balancing performance, practicality, and security in a single framework.

\subsubsection{Plaintext Training Phase}

In the training phase, the convolutional neural network (CNN) described in the preceding sections was trained on the CIFAR-10 dataset, achieving a final classification accuracy of 94.67\%. The dataset was divided into 83.3\% for training (50,000 images) and 16.7\% for testing (10,000 images). During training, the model learned and stored all parameters required for the encrypted inference stage. These parameters include the weights and biases of the convolutional layers, which extract hierarchical features from the input images, and the parameters of the nine batch normalization layers. Each batch normalization layer stores a scale factor (\( \gamma \)) and shift factor (\( \beta \)), along with running estimates of the mean (\( \mu \)) and variance (\( \sigma^2 \)) accumulated during training. An additional small constant (\( \epsilon \)) is used to maintain numerical stability during normalization. For encrypted inference, these batch normalization parameters are not used as separate operations; instead, they are folded into the weights and biases of the fully connected or convolutional layers before encryption, eliminating the need to execute BN steps directly in the encrypted domain and ensuring compatibility with HE constraints. The fully connected layer consist of a layer with 512 units and an output layer with 10 units. The folded weights and biases from the layers is preserved for use during encrypted inference. To reduce the computational complexity under homomorphic encryption, the feature vectors output from the convolutional blocks and batch normalization, just before entering FC1, are pre-computed and stored in plaintext, along with their corresponding ground-truth labels. These pre-computed features are then used as direct inputs to the encrypted inference process, allowing only the fully connected computations to be carried out in the encrypted domain. By retaining the convolutional parameters, batch normalization statistics, folded fully connected weights and biases, pre-computed features, and labels, the system can perform encrypted inference without re-training the model and without recomputing convolutional outputs in ciphertext, thereby reducing multiplicative depth and computational overhead.
\noindent \subsubsection{Encrypted Inference Phase}
The inference phase employs a hybrid design to balance privacy, accuracy, and computational efficiency. While the CKKS scheme enables ciphertext arithmetic, directly evaluating high-degree polynomials under full encryption can quickly consume the noise budget. 
Therefore, in our implementation, the client decrypts each intermediate activation locally, computes the polynomial powers in plaintext, and re-encrypts each term before aggregation. 
This ensures that intermediate values are never exposed to the untrusted server, preserving end-to-end data confidentiality while reducing multiplicative depth and runtime by several orders of magnitude compared to fully homomorphic evaluation. 

\begin{table}[H]
\centering
\caption{CKKS encryption parameters used during inference.}
\label{tab:ckks}
\begin{tabular}{lcccc}
\toprule
Poly. degree & Mod. chain (bits) & Scale & Security & Packing \\
\midrule
8192 & [60, 40, 40, 60] & $2^{40}$ & 128-bit & Disabled ($k{=}1$) \\
\bottomrule
\end{tabular}
\end{table}

In this work, we employ the CKKS homomorphic encryption scheme, as shown in Table IV, to enable privacy-preserving inference on encrypted data. CKKS is a leveled homomorphic encryption scheme designed for approximate arithmetic on real numbers, making it well-suited for deep learning applications where floating-point operations dominate \cite{cheon2017ckks}. In CKKS, vectors are encoded into polynomials over $\mathbb{Z}[X]/(X^N+1)$ and encrypted with a public/secret key pair. Once encrypted, CKKS supports approximate addition and multiplication directly over ciphertexts, enabling the evaluation of linear layers and polynomial activations without decryption. These standard operations are well established in the literature and form the basis for encrypted neural network inference \cite{bian2023heir, Hesamifard2017CryptoDL}.

To eliminate batch normalization (BN) at inference, we fold BN into the preceding layer. If a linear layer produces
\begin{equation}
z = W\mathbf{x} + \mathbf{b}, \tag{6}
\end{equation}
and the subsequent BN is
\begin{equation}
BN(z) = \gamma \odot \frac{z - \mu}{\sqrt{\sigma^2 + \epsilon}} + \beta, \tag{7}
\end{equation}
then there exist folded parameters $W'$ and $\mathbf{b}'$ such that
\begin{equation}
BN(W\mathbf{x} + \mathbf{b}) \equiv W'\mathbf{x} + \mathbf{b}', \tag{8}
\end{equation}
with \begin{equation}
W' = \mathrm{diag}\!\left(\frac{\gamma}{\sqrt{\sigma^2+\epsilon}}\right) W, 
\qquad
\mathbf{b}' = \frac{\gamma}{\sqrt{\sigma^2+\epsilon}} \odot (\mathbf{b}-\mu) + \beta. \tag{9--10}
\end{equation}

In our pipeline, FC1 uses the BN-folded parameters $(W^{(1)'}, \mathbf{b}^{(1)'})$. During inference, the TenSEAL CKKS context is configured in standard-precision mode for computational efficiency. Let
\begin{equation}
\mathbf{x}^{(i)} \in \mathbb{R}^d \tag{11}
\end{equation}
denote the $i$-th precomputed feature vector. In the encrypted domain, this is encoded and encrypted as
\begin{equation}
c_x^{(i)} = \text{Encrypt}_{pk}(\text{Encode}(\mathbf{x}^{(i)})). \tag{12}
\end{equation}

The first fully connected layer has weight matrix
\begin{equation}
W^{(1)} \in \mathbb{R}^{h \times d}, \tag{13}
\end{equation}
and bias vector
\begin{equation}
\mathbf{b}^{(1)} \in \mathbb{R}^h. \tag{14}
\end{equation}
The $j$-th neuron output is
\begin{equation}
z_{1,j}^{(i)} = \langle c_x^{(i)}, W^{(1)}_{j,:} \rangle + b^{(1)}_j. \tag{15}
\end{equation}
Encrypted plaintext dot products are computed using slotwise multiplications and rotations, as supported in TenSEAL \cite{tenseal2021}.

After FC1, the degree-4 polynomial approximation is placed as follows: 
\begin{align}
a_j^{(i)} =& -0.00068481 \,(z_{1,j}^{(i)})^4 
+ 0.0887234775 \,(z_{1,j}^{(i)})^2 \notag \\
&+ 0.5 \, z_{1,j}^{(i)} + 0.738099333. \tag{19}
\end{align}

This polynomial is HE-compatible and avoids costly non-polynomial operations.

The predicted class is obtained as
\begin{equation}
\hat{y}^{(i)} = \arg\max_k \ \ell_k^{(i)}. \tag{23}
\end{equation}
For a test set of size $T$, partitioned into batches $\{B_m\}_{m=1}^M$, the accuracy is
\begin{equation}
\text{Acc} = \frac{1}{T} \sum_{m=1}^M \sum_{i \in B_m} 
\mathbf{1}\left\{\arg\max_k \ \ell_k^{(i)} = y^{(i)} \right\}. \tag{24}
\end{equation}

The encrypted multiplicative depth is dominated by one encrypted plaintext dot product. The polynomial activation is hybrid and thus does not consume encrypted powers. This depth fits comfortably within the noise budget of the standard-precision CKKS setting used here. Running inference on 10,000 samples achieved an accuracy of $94.4\%$ with a total processing time of $2.42$ seconds per sample.

All encrypted experiments were performed with TenSEAL using the CKKS scheme in single-sample mode (\(k{=}1\)) without ciphertext packing. 
The encryption parameters were set to a polynomial modulus degree of 8192, coefficient modulus bit sizes of [60, 40, 40, 60], and a global scale of \(2^{40}\), corresponding to a 128-bit security level. 
This configuration maintained sufficient noise budget for the degree-4 polynomial activation without requiring bootstrapping.

\section{Experimental Setup and Results}

\noindent \textbf{Experimental Setup:}The proposed framework was evaluated using the CIFAR-10 dataset, which contains 60,000 color images across 10 classes with 50,000 training and 10,000 testing. The CNN architecture employed follows the design described in the Methodology, consisting of convolutional blocks with batch normalization folded into the preceding layers, and one fully connected layers, connected via a degree-4 polynomial approximation of the Softplus activation function.
\noindent Encrypted inference was implemented using the CKKS scheme through the TenSEAL library, enabling approximate arithmetic over real numbers with SIMD-style batching. To reduce multiplicative depth, only the fully connected layers and polynomial activation were evaluated in the encrypted domain, while convolutional features were precomputed in plaintext.  
\noindent All experiments were executed using Intel i9-14900HX CPU with 24 cores and 32 GB RAM. Encryption parameters were selected to balance precision and efficiency, ensuring sufficient noise budget for two encrypted–plaintext dot products per inference.
\noindent \textbf{Results:} The proposed framework achieved a classification accuracy of $94.4\%$ on the CIFAR-10 dataset with an inference time of $2.42$ seconds per each sample, as shown in Table V. 

\begin{table}[H]
\centering
\caption{Latency breakdown for one encrypted inference (single sample, no packing).}
\label{tab:latency}
\begin{tabular}{lcc}
\toprule
\textbf{Stage} & \textbf{Time (s)} & \textbf{Share (\%)} \\
\midrule
Client encode \& encrypt   & 0.0363 & 1.5 \\
Fully connected (FC) layer & 1.09   & 45.0 \\
Polynomial activation      & 1.23   & 50.8 \\
Decryption                 & 0.068  & 2.8 \\
\midrule
\textbf{Total per sample}  & \textbf{2.42} & \textbf{100} \\
\bottomrule
\end{tabular}
\end{table}

This represents a notable improvement over \cite{obla}, who reported $89.91`\%$ accuracy under a similar degree-4 polynomial approximation, demonstrating the effectiveness of our activation design and architectural modifications.  
\noindent While higher-degree polynomials may yield closer approximations and potentially higher accuracy, they also increase multiplicative depth and computational cost under homomorphic encryption, requiring either larger noise budgets or costly bootstrapping. Our degree-4 polynomial strikes a practical balance, maintaining high accuracy while ensuring computational feasibility.

\begin{table}[H]
\centering
\caption{Approximation details of activation functions and their corresponding accuracies.}
\begin{tabular}{lcccc}
\toprule 
\makecell{\textbf{Activation} \\ \textbf{Function}} & 
\makecell{\textbf{Approximation} \\ \textbf{Degree}} & 
\makecell{\textbf{RAM and} \\ \textbf{Processing}} & 
\textbf{Accuracy} \\
\midrule
Softplus & 4 & 
\makecell{
Intel\textsuperscript{\textregistered} \\
Core\textsuperscript{TM} \\
i9-14900HX \\
(24 cores / 32 threads, \\
2.20~GHz) \\
with 31.71~GB RAM
} & 94.4\% \\

Softplus \cite{obla} & 4 &
\makecell{
Xeon Silver 4114 \\
CPU @ 2.20~GHz \\
with 192~GB RAM
} & 89.91\% \\
\bottomrule
\end{tabular}
\end{table}

\noindent In terms of architecture, deeper convolutional networks are capable of capturing richer hierarchical representations but incur significant computational overhead when evaluated under homomorphic encryption. Each additional layer increases the multiplicative depth, ciphertext noise, and number of bootstrapping operations required, all of which amplify runtime costs. The proposed framework establishes a balanced compromise between representational complexity and cryptographic efficiency, being sufficiently expressive to achieve competitive accuracy on CIFAR-10 while maintaining encrypted inference within a practical latency range. This design choice enables meaningful encrypted learning without the infeasibility associated with fully deep architectures in homomorphic settings.

\subsection*{Evaluation on CIFAR-100}
To further evaluate the generalizability of the proposed polynomial activation function and encryption-aware architecture, we extended the experiments to the CIFAR-100 dataset using the same PreAct-ResNet-18 architecture previously described. The standard plaintext baseline for ResNet-18 on CIFAR-100 achieves approximately 79--80\% under standard augmentation and training configurations \cite{resnet18_cifar100}. For encrypted inference, the trained model weights were evaluated under the CKKS homomorphic encryption scheme with parameters configured as follows: polynomial modulus degree $n = 16{,}384$, coefficient modulus bit sizes $[60, 40, 60]$, and global scale $2^{40}$ as shown in Table VII. To account for the broader activation distribution in CIFAR-100, the polynomial activation was re-fit to the model’s actual logit range to maintain numerical precision under encryption. Using this configuration, the encrypted inference achieved 78.16\% accuracy, which is nearly identical to the plaintext result. This demonstrates that the proposed activation and framework preserve both numerical fidelity and predictive performance even when scaling from smaller datasets such as CIFAR-10 to more complex, higher-dimensional benchmarks like CIFAR-100.

\begin{table}[h!]
\centering
\caption{CKKS Parameters Used for Encrypted Inference on CIFAR-100}
\label{tab:ckks_params}
\begin{tabular}{l c}
\hline
\textbf{Parameter} & \textbf{Value} \\ 
\hline
Polynomial Modulus Degree ($n$) & 16,384 \\ 
Coefficient Modulus Bit Sizes & [60, 40, 60] \\ 
Global Scale & $2^{40}$ \\ 
Security Level & $\approx$ 128-bit \\ 
Evaluation Depth & 4 (Post-Activation) \\ 
\hline
\end{tabular}
\end{table}

\section{Acknowledgement}

This work is supported by the National Science
Foundation (NSF) Award number 2409093.

\bibliographystyle{IEEEtran}
\bibliography{references}

\end{document}